\documentclass[,draft]
  {aipproc}

\layoutstyle{6x9}

\begin{document}

\title{On the Dynamical Stability of $\gamma$ Cephei,
\hskip 1in
an S-Type Binary Planetary System}

\author{Nader Haghighipour}{
  address={Department of Terrestrial Magnetism and 
NASA Astrobiology Institute, \\Carnegie Institution of
Washington, 5241 Broad Branch Road, Washington, DC 20015}
}

\begin{abstract}
Precision radial velocity measurements of the $\gamma$ Cephei
(HR8974) binary system suggest the existence of a
planetary companion with a minimum mass of 1.7 Jupiter-mass
on an elliptical orbit with a $\sim$2.14 AU semimajor axis and 0.12
eccentricity \cite{hatzes03}. I present in this paper a summary of
the results of an extensive numerical study of the
orbital stability of this three-body system for different
values of the semimajor axis and orbital eccentricity of the
binary, and also the orbital inclination of the planet.
Numerical integrations indicate that the system is stable
for the planet's orbital inclination ranging from 0 to
60 degrees, and for the binary's orbital eccentricity
less than 0.5. The results also indicate that for large
values of the inclination, the system may be locked in a
Kozai resonance. 
\end{abstract}

\maketitle


\section{INTRODUCTION}

The existence of planets in binary star systems is no longer 
a mere idea. Approximately 35\% of extrasolar planets
discovered till 2002 exist in multiple star systems
\cite{kortenkamp01}. These systems are mostly wide binaries
with separations between 500 and 750 AU, and with planets
revolving around one of the stars. At such large distances,
the perturbative effect of one star on the formation
and dynamical evolution of planets around the other star
is entirely negligible. A recently discovered Jupiter-like
planet around the primary of the $\gamma$ Cephei binary
system is, however, an exception to this rule.

Gamma Cephei is a close spectroscopic binary composed of a 
1.59 solar-mass K1 IV subgiant and a 0.4 solar-mass
red M dwarf \cite{fur03}. The semimajor axis of this
system has been reported to have a lower value of 18.5 AU
\cite{hatzes03} and an upper value of 21 AU \cite{griffin02}.
Precision radial velocity measurements suggest that a
planet with a minimum mass of 1.7 Jupiter-mass revolves around the
primary of this system on an elliptical orbit with an
eccentricity of 0.12 $\pm$ 0.05 and a semimajor axis of
approximately 2.14 AU. Being the first discovered S-type
binary-planetary system\footnote{As classified by Rabl $\&$ Dvorak (1988),
a binary-planetary system is called S-type when the planet
revolves around one of the stars, and P-type when the planet revolves
around the entire binary.}, it is quite valuable to investigate
whether this system is dynamically stable,
and for what values of its orbital parameters its stability will
remain. In this paper I present a summary of the results of an
extensive numerical study of the dynamical stability of this
system for different values of the orbital parameters of the binary
and also the orbital inclination of the planet.
A more comprehensive study of the dynamics of this system
are to be published elsewhere.

\section{NUMERICAL ANALYSIS}
\label{system}

An important quantity in determining the stability of a 
planet's orbit in a binary system
is its semimajor axis. \citet{Dvorak88a} and 
\citet{holman99} have presented an empirical formula for
the maximum value of the semimajor axis of a stable planetary
orbit in S-type binary-planetary systems. 
The value of the {\it critical semimajor axis}, $a_c$ is given by
\begin{eqnarray}
&{{a_c}/{a_b}}=(0.464\pm 0.006)+
(-0.380 \pm 0.010)\mu + (-0.631\pm0.034) {e_b}\nonumber\\
&+ (0.586 \pm 0.061) \mu {e_b} + (0.150 \pm 0.041) {e_b^2} 
+(-0.198 \pm 0.047)\mu {e_b^2}
\end{eqnarray}
\noindent
where $a_b$ is the semimajor axis of the binary, and 
$\mu={m_2}/({m_1}+{m_2})$ and $e_b$ represent the mass-ratio 
and orbital eccentricity of the binary, respectively. 
In the definition of the mass-ratio $\mu$, 
$m_1$ and $m_2$ are the masses of the primary
and secondary stars. To study to what extent equation (1) can be
applied to the orbital stability of $\gamma$ Cephei,
this system was numerically integrated using a conventional
Bulirsch-Stoer integrator. Numerical integrations were carried out for
values of $e_b$ ranging from 0.25 to 0.65 with steps of 0.05.
When considered coplanar,
the system was stable for all the values of ${e_b}<0.5$
at all times. However, for the value of the binary eccentricity
larger than 0.5, the system became unstable in less than 1000 years.
Figure 1 shows the semimajor axes and eccentricities of the system
for two cases of ${e_b}=0.25$ and 0.45. 

The stability of the system was also studied for different values
of the orbital inclination of the planet. For each above-mentioned
value of the binary eccentricity, the initial inclination of the orbit
of the planet with respect to the plane of the binary was
chosen from the values of 1, 5, 10, 20, 40, and 60 degrees.
For values of ${e_b}<0.5$, 
the system was stable for all inclinations less than
60 degrees.
Figure 3 shows the results of sample runs for the values
of the planet's orbital inclination equal to 5, 10, and 20 
degrees.  Also, as expected, in some cases of large inclinations and
for eccentric binaries, the planet was locked in a Kozai
resonance (Fig. 3).


\begin{theacknowledgments}
This work is partially supported by the NASA Origins of 
the Solar System Program under
grant NAG5-11569, and also the NASA Astrobiology Institute
under Cooperative Agreement NCC2-1056.
\end{theacknowledgments}

\bibliographystyle{aipproc}

\begin{thebibliography}
\expandafter\ifx\csname natexlab\endcsname\relax\def\natexlab#1{#1}\fi
\providecommand{\enquote}[1]{``#1''}
\expandafter\ifx\csname url\endcsname\relax
  \def\url#1{\texttt{#1}}\fi
\expandafter\ifx\csname urlprefix\endcsname\relax\def\urlprefix{URL }\fi

\bibitem[Hatzes et al. (2003)]{hatzes03}
Hatzes, A. P., Cochran, W. D., Endl, M., McArthur, B.,
Paulson, D. B., Walker, G. A. H., Campbell, B., and Yang, S.,
to appear in \emph{Astrophys. J.}, \textbf{599}, 1383--1394 (2003).
\bibitem[Kortenkamp, Wetherill, $\&$ Inaba (2001)]{kortenkamp01}
Kortenkamp, S. J., Wetherill, G. W., and Inaba, S.,
\emph{Science}, \textbf{293}, 1127--1129 (2001).
\bibitem[Fuhrmann (2003)]{fur03}
Fuhrmann, K., \emph{Astron.Nachr.}, in press (2004).
\bibitem[Griffin, Carquillat, $\&$ Ginestet (2002)]{griffin02}
Griffin, R. F., Carquillat, J. M, and Ginestet, N.,
\emph{The Observatory}, \textbf{122}, 90--109 (2002).
\bibitem[Rabl $\&$ Dvorak (1988)]{Dvorak88a}
Rabl, G., and Dvorak, R., \emph{Astron. Astrophys.}, 
\textbf{191}, 385--391 (1988).
\bibitem[Holman $\&$ Wiegert (1999)]{holman99}
Holman, M. J., and Wiegert, P. A., \emph{Astron. J.}, 
\textbf{117}, 621--628 (1999).


\end{thebibliography}

\vfill
\eject

\begin{figure}
\caption{See Fig1.gif
}
\end{figure}

\clearpage

\begin{figure}[b]
%
  \caption{See Fig2.gif
}
  \hfill
\end{figure}

\clearpage

\begin{figure}[b]
%
  \caption{See Fig3.gif
}
  \hfill
\end{figure}

\end{document}